\documentclass[11pt]{article}
\usepackage{jheppub}

\usepackage{epsfig}
\usepackage{graphicx}
\def\be{\begin{equation}}
\def\ee{\end{equation}}
\def\bea{\begin{array}}
\def\eea{\end{array}}
\def\beqa{\begin{eqnarray}}
\def\eeqa{\end{eqnarray}}
\def\beqas{\begin{eqnarray*}}
\def\eeqas{\end{eqnarray*}}

\def\bp{\begin{picture}}
\def\ep{\end{picture}}
\def\bc{\begin{center}}
\def\ec{\end{center}}
\def\bfig{\begin{figure}}
\def\efig{\end{figure}}

\def\bit{\begin{itemize}}
\def\eit{\end{itemize}}
\def\nn{\nonumber}
\def\f{\frac}

\def\[{\left[}
\def\]{\right]}
\def\({\left(}
\def\){\right)}

\def\..{\left.}
\def\.{\right.}
\def\tl{\tilde}
\def\ra{\rightarrow}
\def\la{\leftarrow}

\def\NPB#1,{{ Nucl.\ Phys.\ B }{\bf #1},}
\def\PLB#1,{{ Phys.\ Lett.\ B }{\bf #1},}
\def\EPJC#1,{{ Eur.\ Phys.\ Jour.\ C }{\bf #1},}
\def\PRD#1,{{ Phys.\ Rev.\ D }{\bf #1},}
\def\PRL#1,{{ Phys.\ Rev.\ Lett.\ }{\bf #1},}
\def\MPLA#1,{{Mod.\ Phys.\ Lett.\ A }{\bf #1},}

\def\da{\dagger}

\def\la{\lambda}

\def\al{\alpha}
\def\bt{\beta}
\def\ka{\kappa}

\def\ep{\epsilon}

\def\pr{\prime}

\begin{document}

\title{A split SUSY model from SUSY GUT}

\author[a,b]{Fei Wang,}
\emailAdd{feiwang@zzu.edu.cn}

\author[c]{Wenyu Wang,}
\emailAdd{wywang@mail.itp.ac.cn}

\author[b]{Jin Min Yang}
\emailAdd{jmyang@itp.ac.cn}

\affiliation[a]{Department of Physics and Engineering, Zhengzhou University,
Zhengzhou 450000, P. R. China}

\affiliation[b]{State Key Laboratory of Theoretical Physics, Institute of Theoretical Physics,
Chinese Academy of Sciences, Beijing 100190, P. R. China}

\affiliation[c]{Institute of Theoretical Physics, College of Applied Science,
Beijing University of Technology,
Beijing 100124, P. R. China}

\abstract{
We propose to split the sparticle spectrum from
the hierarchy between the GUT scale and the Planck scale.
A split supersymmetric model,
which gives non-universal gaugino masses, is built with proper
high dimensional operators in the framework of SO(10) GUT.
Based on a calculation of two-loop beta functions for gauge couplings
(taking into account all weak scale threshold corrections), we check the gauge coupling unification
and dark matter constraints (relic density and direct detections).
We find that our scenario can achieve the gauge coupling unification and satisfy
the dark matter constraints in some part of parameter space.
We also examine the sensitivity of the future XENON1T experiment
and find that the currently allowed parameter
space  in our scenario
can be covered for a neutralino dark matter below about 1.0 TeV.
}

\maketitle

\section{Introduction}
The discovery of a 125 GeV Higgs boson \cite{atlas,cms} by both the
ATLAS and CMS collaborations has completed the Standard Model (SM).
The experimental data reported so far by LHC agree quite well with the SM predictions.
On the other hand, the lack of significant hints in electroweak precision
tests and the absence of evidences for new particle contents at the LHC
challenge many proposals of new physics beyond the SM,
including weak scale supersymmetry (SUSY).

SUSY, which was regarded for a long time as one of
the most appealing extensions of the SM, has many desirable features.
For example, the observed 125 GeV Higgs boson falls within the narrow
$115-135$ GeV window predicted by the minimal supersymmetric standard
model (MSSM). Besides, the genuine unification of gauge
couplings \cite{su5,gutmssm,Einhorn:1981sx,Marciano:1981un},
which can hardly be achieved in the SM, can be successfully realized
in the framework of low energy SUSY.
Also, the puzzle of cosmic dark matter can naturally be explained in SUSY.

Although SUSY is appealing, null search results of sparticles at LHC suggest
that either low energy SUSY needs to be tuned or sparticles are well above
the weak scale. In fact, the LHC data has already set a limit \cite{cmssm1,cmssm2}
$m_{\tilde g} > 1.5$ TeV for $m_{\tl{q}} \sim m_{\tl{g}}$ and $m_{\tl{g}}\gtrsim 1$ TeV
for $m_{\tl{q}} \gg m_{\tl{g}}$ for certain popular CMSSM models.
On the other hand, the observed mass of the Higgs boson
requires rather large loop effects of top squarks in the MSSM and CMSSM,
which implies some extent of fine-tuning \cite{cao}.
So naturalness in SUSY may be realized in a more involved way even
though it was initially proposed to solve the hierarchy problem.

Split SUSY, proposed in \cite{nima,giudice1,giudicenima},
gives up the naturalness criterion while keeps the other two main advantages:
the gauge coupling unification and viable dark matter candidates.
This scenario assumes a very high scalar mass scale $M_S$ and
the low energy spectrum contains only the gauginos
and higgsinos as well as a fine-tuned Higgs boson in addition to the
SM sector.
The SUSY CP and flavor problems can naturally be
solved in this scenario due to very heavy sfermions.
The latest results of the Higgs mass from the LHC measurement,
together with the requirement of
gauge coupling unification, suggest a scalar superpartner
mass scale roughly of order $M_S\sim 100-1000$ TeV \cite{giudice2,minisplit,FWY},
which indicates a moderately split spectrum for split SUSY.
The hierarchy of sparticle spectrum is argued to be determined by the gauge
loop factor \cite{minisplit}.
We propose in this paper to split the sparticle spectrum from
the hierarchy between the GUT scale and the Planck scale.

Actually, there are various ways to split the sparticle spectrum.
In this work we propose to use the generalized gravity
mediation \cite{msugra,Li:2010xr,fei} with non-renormalizable
K\"ahler potential and superpotential.
Because GUT is one of the retained motivations of split SUSY,
certain high-representation Higgs fields of GUT group could appear
in the non-renormalizable K\"ahler potential and superpotential.
Such high-representation fields could not only amend the gauge coupling
unification condition at the GUT scale,
but also establish new relations among various theory inputs.
In our scenario we have non-universal gaugino masses at the GUT scale
(for other scenarios giving non-universal gaugino masses,
see \cite{ngaugino1,ngaugino2,ngaugino}).
Such non-universal gaugino masses can naturally appear
with non-renormalizable K\"ahler potential and superpotential involving various
high-representation Higgs fields.

This paper is organized as follows. In Sec. \ref{model} we present our model.
In Sec. \ref{dm} we check the phenomenology of our model.
Based on a calculation of two-loop beta functions for gauge couplings
(taking into account all weak scale threshold corrections), we check
the gauge coupling unification and the dark matter constraints.
The future XENON1T sensitivity to our scenario is also examined.
Finally, in Sec. \ref{conclusion} we give our conclusions.

\section{A split SUSY model from SUSY GUT}\label{model}
There are many possible ways to mediate the SUSY breaking
effects from the hidden sector to the visible sector.
A very interesting and predictive possibility is
the gravity mediation. With certain non-renormalizable terms, proper
soft SUSY breaking parameters can be generated. In many popular
gravity mediation scenarios, the K\"ahler potential
is assumed to be minimal. However, a general K\"ahler potential
seems to be more natural. When certain high-representation chiral
fields for the GUT group are involved in the non-renormalizable K\"ahler potential,
the kinetic terms of superfields could have other contributions
after the GUT symmetry breaking. New non-renormalizable terms in the
superpotential involving high-representation fields could also be important.

In general, the non-vanishing F-term VEVs of certain fields which break
SUSY could be either gauge singlets or non-singlets. In this section,
we propose that SUSY breaking is triggered by the GUT group non-singlet
F-term VEVs. The gaugino and sfermion masses will be generated by
some non-renormalizable operators which could arise from integrating
out certain gravitational effects. In order to get more compact spectrum
and simplify the relevant expressions, we adopt the SO(10) GUT group with the
Georgi-Glashw SU(5) GUT group as an intermediate stage in the symmetry breaking chain:
\beqas
SO(10){\f{}{~\bf 16,\overline{16}~~}} SU(5) \times U(1)_X {\f{}{\bf ~~54 ({\bf 45})~~}}
SU(3)_c\times SU(2)_L\times U(1)_Y.
\eeqas
Here $U(1)_X$ is broken by possible $\nu_L^c$ component VEVs of
$H_{\bf 16}$ and $\bar{H}_{\overline{\bf 16}}$ Higgs fields.
Therefore, most of the results in this paper are also valid in SU(5) GUT.
The possibility of SO(10) to Pati-Salam will be considered elsewhere.

In order to accommodate the non-minimal K\"ahler potential, we need to know
the group products of various SO(10) representations \cite{slansky}.
The spinor representation (for the matter part) can be decomposed as
 \beqa
 \overline{\bf 16} \otimes {\bf 16}={\bf 1}\oplus {\bf 45}\oplus{\bf 210},
 \eeqa
while the adjoint  representation for gaugino is
\beqa
({\bf 45}\otimes{\bf 45})_{\rm symmetric}={\bf 1}
\oplus {\bf 54}\oplus {\bf 210}\oplus{\bf 770}.
\eeqa
and the fundamental representation of Higgs is
\beqa
{\bf 10}\otimes {\bf 10}={\bf 1}\oplus{\bf 45}\oplus{\bf 54}.
\eeqa
We assume the non-minimal gauge kinetic term for vector supermultiplets
\beqa
{\cal L}=\int d^2\theta  W^\al(\delta_{\al\beta}+\eta\f{ \Phi_{\al \bt}}{M_{*}})W^\bt~.
\eeqa
with $M_*$ being the reduced Planck scale.
So the kinetic part for gauge field with non-minimal K\"ahler potential is given by
 \beqa
 {\cal L}=-\f{1}{4 k} Tr [F_{\mu\nu} F^{\mu\nu}]-\f{\eta}{4 k M_{*}} Tr[F_{\mu\nu} \Phi F^{\mu\nu}]~,
 \eeqa
 with $k$ being the normalization factor for various representations
according to $Tr(T^aT^b)=k\delta^{ab}$.
After the GUT non-singlet develops a VEV, $\langle\Phi\rangle=v+F_\Phi \theta^2$
with $v^2\gg F_\Phi$,  the unification condition turns into
\beqa
\label{nGUT}
g_1^2(M_X)(1+\f{\eta v}{M_{*}}\delta_1)=g_2^2(M_X)(1+\f{\eta v}
{M_{*}}\delta_2)=g_3^2(M_X)(1+\f{\eta v}{M_{*}}\delta_3)
\eeqa
with $\delta_{3,2,1}$ being the appropriate group factors
for $SU(3)_c,SU(2)_L,U(1)_Y$, respectively.
If the hierarchy between the (first step Georgi-Glashow SU(5)) GUT scale
and the Planck scale is not small, the previous GUT conditions turn into
two independent new GUT conditions:
\beqa
F_1\equiv\f{\f{g_1^2}{g_2^2}(M_X)-1}{\f{g_1^2}{g_3^2}(M_X)-1}=\f{\delta_2-\delta_1}{\delta_3-\delta_1}=-\f{2}{3}~,\\
F_2\equiv\f{\f{g_2^2}{g_1^2}(M_X)-1}{\f{g_2^2}{g_3^2}(M_X)-1}=\f{\delta_1-\delta_2}{\delta_3-\delta_2}=~\f{2}{5}~.
\eeqa
For non-singlet $\Phi$, the F-term $F_\Phi$ can be decomposed
as $(F_\Phi)_{ab}=F_U\cdot A_{ab}$  with $A_{ab}$ being the
group factor and $F_U$ the universal part.
The review of the group structure can be found in \cite{slansky}.

The group structure of the {\bf 24} component F-term VEV of
${\bf 54}$ representation Higgs can be written in terms of $10\times 10$ matrix
\beqa
\langle F_{\bf 54}\rangle_{ab}=F_{\bf 54}^U A_{ab}=F_{\bf 54}^U\sqrt{\f{3}{5}}
(~\f{1}{3},~\f{1}{3},~\f{1}{3},-\f{1}{2},-\f{1}{2},~\f{1}{3},
~\f{1}{3},~\f{1}{3},-\f{1}{2},-\f{1}{2}).
\eeqa
The gaugino will get contribution from
\beqa
{\cal L}\supseteq \f{<F_{\Phi}>_{\al\beta} }{M_{*}}\la^\al\la^\beta
\eeqa
with the F-term VEVs $(F_{\bf 54})_{ab}=F_{\bf 54}^U.A_{ab}$. Here the
universal part $F_{\bf 54}^U$ is independent of the group structure.
Then the non-universal gaugino masses are given by
\beqa
M_1=- \f{1}{6}\sqrt{\f{3}{5}}m_{1/2},~~
M_2=-\f{1}{2}\sqrt{\f{3}{5}}m_{1/2},~~
M_3=\f{1}{3}\sqrt{\f{3}{5}}m_{1/2}
\eeqa
with
\beqa
m_{1/2}=\f{F_{\bf 54}^U}{M_*}.
\eeqa
The sfermion masses and kinetic term will be generated by the following
non-renormalizable K\"ahler potential
\beqa
K=\f{1}{M_*^2}\phi_a^\da (\Phi^\da\Phi)_{ab}\phi_b
\eeqa
with proper F-term and lowest component VEVs of $\Phi$, respectively.
Here we assume that the universal part of the kinetic terms $\Phi^\da\Phi$ is approximately canceled by a similar high dimensional operators with a lowest component VEV of singlet.

We know from the group theory that possible contributions to K\"ahler
potential for matter content (filled in {\bf 16} representation of SO(10))
can arise from the following type of Higgs fields
\beqa
{\bf 1}\oplus{\bf 45}\supset{\bf 54}\otimes {\bf 54}~,~~~~~
{\bf 1}\oplus{\bf 210}\supset \overline{\bf 16} \otimes {\bf 16}~.
\eeqa
So the F-term VEV of the {\bf 54} representation chiral superfield
will contribute to non-universal sfermion masses of order $({F_{\bf 54}}/{M_*})^2$
in addition to possible contributions from $F_{\bf 16}$.
We assume $F_{\bf 16}\sim F_{\bf 54}$ and we always have $F_{\bf \overline{16}}=F_{\bf 16}$.
Since $v_{\bf 16}\gg v_{\bf 54}$, the {\bf 54} representation field will give
sub-leading contributions
and we will not include them explicitly in the following expressions.
From the group structure there are several possible
contractions for matter fields $\phi$ with the form of K\"ahler potential
\small
\beqa
&&K\supset\f{1}{M_*^2}\sum\limits_{a=1}^3(\phi^\da_{a,\bf 16}\otimes
\phi_{a,\bf 16})^{\bf 1}[(d_{\bf 1}\Phi^\da_{H_{\bf {16}}}+
\tl{d}_{\bf 1}\Phi_{\bar{H}_{\bf \overline{16}}})\otimes (f_{\bf 1}\Phi_{H_{\bf 16}}
+\tl{f}_{\bf 1}\Phi^\da_{\bar{H}_{\bf \overline{16}}})]^{\bf 1}~~~\nn\\
&&+\f{1}{M_*^2}\sum\limits_{a=1}^3(\phi_{a,\bf 16}^\da\otimes\phi_{a,\bf 16})^{\bf 45}_{mn}
[(d_{\bf 45}\Phi^\da_{H_{\bf {16}}}+\tl{d}_{\bf 45}\Phi_{\bar{H}_{\bf \overline{16}}})
\otimes (f_{\bf 45}\Phi_{H_{\bf 16}}
+\tl{f}_{\bf 45}\Phi^\da_{\bar{H}_{\bf \overline{16}}})]^{\bf 45}_{mn}~~\nn\\
&&+\f{1}{M_*^2}\sum\limits_{a=1}^3(\phi^\da_{a,\bf 16}\otimes\phi_{a,\bf 16})^{\bf 210}_{mnlp}
[(d_{\bf 210}\Phi^\da_{H_{\bf {16}}}+\tl{d}_{\bf 210}\Phi_{\bar{H}_{\bf \overline{16}}})
\otimes (f_{\bf 210}\Phi_{H_{\bf 16}}
+\tl{f}_{\bf 210}\Phi^\da_{\bar{H}_{\bf \overline{16}}})]^{\bf 210}_{mnlp}
\label{eq215}
\eeqa
\normalsize
where $d,f,\tl{d},\tl{f}$ denote the corresponding combination
coefficients and  $a$ is the family index.

It can be checked that only the first term in Eq. (\ref{eq215})
contributes in this scenario.
Then the soft sfermion masses are given by
\beqa
\epsilon^2\tl{m}_{\bf 16}^2=d_{\bf 1}f_{\bf 1}\f{F_{\bf 16}^*F_{\bf 16}}{M_*^2}
+\tl{d}_{\bf 1}\tl{f}_{\bf 1}\f{F_{\bf \overline{16}}^*F_{\bf \overline{16}}}{M_*^2}
+d_{\bf 1}\tl{f}_{\bf 1}\f{F_{\bf 16}^*F_{\bf \overline{16}}}{M_*^2}
+\tl{d}_{\bf 1}f_{\bf 1}\f{F_{\bf \overline{16}}^*F_{\bf 16}}{M_*^2}
\eeqa
after we take into account the normalization factor $\epsilon^2=v^2_{\bf 16}/M_*^2$ of the kinetic term.
So we have
\beqa
\tl{m}_{\bf 16}^2\sim \f{F_{\bf 16}^*F_{\bf 16}}{M_*^2 \epsilon^2}\sim \f{F_{\bf 16}^*F_{\bf 16}}{v_{\bf 16}^2}.
\eeqa
We can see that the typical gaugino mass scale $m_{1/2}$ is
suppressed by a factor $\epsilon$  relative to the sfermion mass scale.

The soft SUSY breaking masses for the Higgs potential can be similarly obtained
\beqa
\label{shiggs}
K&\supset&\f{1}{M_*^2}(\phi^\da_{\bf 10}\otimes \phi_{\bf 10})^{\bf 1}
[(g_{\bf 1}\Phi^\da_{H_{\bf {16}}}+\tl{g}_{\bf 1}
\Phi_{\bar{H}_{\bf \overline{16}}})\otimes (h_{\bf 1}\Phi_{H_{\bf 16}}
+\tl{h}_{\bf 1}\Phi^\da_{\bar{H}_{\bf \overline{16}}})]^{\bf 1}~~~\nn \\
&+&\f{1}{M_*^2}(\phi_{\bf 10}^\da\otimes\phi_{\bf 10})^{\bf 45}_{mn}
[(g_{\bf 45}\Phi^\da_{H_{\bf {16}}}+\tl{g}_{\bf 45}\Phi_{\bar{H}_{\bf \overline{16}}})
\otimes (h_{\bf 45}\Phi_{H_{\bf 16}}
+\tl{h}_{\bf 45}\Phi^\da_{\bar{H}_{\bf \overline{16}}})]^{\bf 45}_{mn},
\eeqa
with also contributions from $F_{\bf 54}$ after we add similar terms involving the
{\bf 54} representation Higgs fields.
Both contributions are at the same order and the soft SUSY breaking
Higgs masses are given by $m_{{H}_{u,d}}^2\sim (m_{1/2})^2$.
Thus we can see that the soft SUSY breaking Higgs mass parameters
can be at the same order as the gaugino.

The trilinear terms will also get contributions from both
${\bf 16}$ and $\overline{\bf 16}$ representation Higgs fields.
The relevant non-renormalizable superpotential has the form
\beqa
W & \supset&  {y_{yukawa}\over {M^2_*}}(\Phi_{\overline{H}_{\bf \overline{16}}}
\otimes \Phi_{H_{\bf 16}})^{\bf 1}
[ C^{\bf 1} ({\bf 16_i\otimes 16_i})_{\bf
10}^m {\bf 10}^m] ~~\nn\\
&+&{y_{yukawa}\over {M^2_*}}(\Phi_{\overline{H}_{\bf \overline{16}}}
\otimes \Phi_{H_{\bf 16}})_{mn}^{\bf 45}
[ C^{\bf 45} ({\bf 16_i\otimes 16_i})_{\bf
120}^{mnl} {\bf 10}^l] ~~\nn\\
&+&{y_{yukawa}\over {M^2_*}}(\Phi_{\overline{H}_{\bf \overline{16}}}
\otimes \Phi_{H_{\bf 16}})_{mnlp}^{\bf 210}
[ C^{\bf 210} ({\bf 16_i\otimes 16_i})_{\bf
126}^{mnlpq} {\bf 10}^q] ~.
\eeqa
Again, here only the first term contributes. From the lowest component
VEV and F-term VEV of $H_{\bf 16}$ and $H_{\overline{\bf 16}}$,
we can obtain the trilinear coupling as
\beqa
\epsilon^2 A_y= C^{\bf 1}[\f{v_{\bf 16}F_{\bf \overline{16}}}{M_*^2}
+\f{v_{\bf \overline{16}}F_{\bf {16}}}{M_*^2}]y_{yukawa}~.
\eeqa
after we normalize the kinetic term for the ${\bf 16}$ representation matter contents.
Thus, we see that
\beqa
A_0=\f{A_y}{y_{yukawa}}\sim \f{F_{\bf \overline{16}}}{v_{\bf 16}}
\eeqa
is typically of the same scale as sfermion masses.

The $B_\mu$-term and $\mu$ term are given by
\beqa
W&\supset& (M+\Phi_{\bf 45}) \phi_{\bf 10}\phi_{\bf 10}+k_1(M+\Phi_{\bf 45})
\f{(\Phi_{\bf 54})}{M_*}\phi_{\bf 10}\phi_{\bf 10}~.
\eeqa
We will not discuss in detail the doublet-triplet splitting problem in GUT and
just use the fact that the combination $M+\langle\Phi_{\bf 45}\rangle$ will lead to
(by possible tuning or some mechanism) proper low energy effective $\mu$ term.

The $B_\mu$-term will be generated after $\Phi_{\bf 54}$ acquires F-term VEVs
\beqa
\label{bmu}
B_\mu=-k_1\sqrt{\f{3}{5}}\f{F_{\bf 54}^U}{M_*}\mu~.
\eeqa
The electroweak symmetry breaking condition can relate different parameters as
\beqa
|\mu|^2&=&-\f{M_Z^2}{2}+\f{1}{\tan^2\beta-1}(m_{H_d}^2-\tan^2\beta m_{H_u}^2)~,\\
2B_\mu&=&\sin2\beta(m_{H_d}^2+ m_{H_u}^2+2|\mu|^2)~.
\eeqa
Although there are many possibilities for the choice of $\mu$ from electroweak
symmetry breaking condition, the relation Eq.(\ref{bmu}) will further constrain
the choice of $\mu$.
Combining Eq.(\ref{bmu}) with the (RGE modified) approximation
$m_{H_d}^2\gtrsim m_{H_u}^2$,
we can see that the symmetry breaking condition requires
$|\mu|^2\sim -m_{H_u}^2\sim B_\mu\sim m_{1/2}^2$
for a negative $m_{H_u}^2$. For a positive $m_{H_u}^2$, it is difficult to reconcile
a large $\tan\beta$ and $m_{H_d}^2\gtrsim m_{H_u}^2$ in case of $\mu\ll m_{h_u}^2$.
Only if the RGE running
can affect greatly the GUT scale relation $m_{H_d}^2= m_{H_u}^2$ can a relatively
heavy $\mu$ be possible.
The cases of $\tan\beta\sim 1$ which indicate $|\mu|^2\gg  |m_{h_d}^2|$ for
both signs of $m_{H_u}^2$ can hardly be compatible with Eq.(\ref{bmu}).

So we see that in our scenario the hierarchy between the GUT scale and the Planck scale
is used in splitting the SUSY soft spectrum in contrast to the
gauge loop factor appeared in simply unnatural supersymmetry  \cite{minisplit}.

\section{Gauge coupling unification and dark matter constraints }\label{dm}
\subsection{Inputs of our model}
In this section we check the gauge coupling unification and dark matter constraints
in our scenario.
From the previous section, we can see the inputs of our scenario at the GUT scale:
\bit
\item The gaugino masses with the raito
\beqa
\label{mratio}
M_1:M_2:M_3=-1:-3:~2\nn
\eeqa
\item The hierarchy between the GUT scale
      (SO(10) breaking scale) and the Planck scale $\epsilon$.
\item The higgsino mass $\mu$ which should be at the same sacle as gaugino mass.
\item The parameter $\tan\beta$. We scan it in the range $1\sim 50$.
\eit
The sfermion mass $m_0$, determined by $m_{1/2}/\epsilon$, and
the trilinear term $A_0\simeq m_0$ are not independent parameters in our model.
This scenario predicts different parameter values in comparison to the universal
gaugino mass scenario \cite{FWY}.
It is well known that the ratios of gaugino masses and the corresponding gauge couplings
in SUSY are RGE-invariant up to one-loop level
  \beqa
  \label{ratio}
  \f{d}{d\ln\mu}\(\f{M_i}{g_i^2}\)=0.
  \eeqa
If we assume universal gaugino masses $M_1=M_2=M_3\equiv M_U$ at the GUT scale,
from the RGE-invariant ratio
 \beqa
 \f{M_1}{g_1^2}=\f{M_2}{g_2^2}=\f{M_3}{g_3^2}=\f{M_U}{g_U^2} \label{universal}
 \eeqa
 and the electroweak scale inputs we can have a mass relation given by
 \beqa
 M_1 : M_2 : M_3\simeq 1: 2 : 6   \label{universal-ew}
 \eeqa
at electroweak scale.
 On the other hand, if we assume non-universal gaugino masses at the GUT scale,
the above gaugino mass ratios at the electroweak scale will be approximately
changed to
 \beqa
 M_1 : M_2 : M_3\simeq 1: 6 : -12 \,. \label{nonuni-ew}
 \eeqa
We should note again that the formula Eq.(\ref{ratio})
can no longer be valid below $M_S$ where heavy scalars decouple.
So the exact gaugino mass ratio in split SUSY should
be obtained by the subsequent RGE running from $M_S$ to EW scale.

\subsection{The gauge coupling unification requirement}
We study the gauge coupling unification with the
two-loop RGE running of gauge couplings,
taking into account all weak scale threshold corrections.
The relevant analytical results for the two-loop beta functions are given in
the appendix. It is well known
that the two-loop RGE running for gauge couplings are scheme independent \cite{martin},
so we use the $\overline{MS}$ couplings in our study of the gauge coupling unification.
 Lacking the full knowledge responsible for GUT symmetry breaking, especially the mechanism used to solve the doublet-triplet splitting problem which is model dependent, we therefore neglect such GUT scale threshold corrections in our study.
  In order to make our calculation reliable,
the first step SU(5) GUT scale must be much lower than the Planck scale so that
the gravitational effects can be neglected. Besides, the GUT scale can not
be too low, otherwise it will lead to fast proton decay.
The constraint is \cite{protondecay,protondecay2,FWY}
\beqa
\label{guttime}
1.0\times 10^{19}{\rm GeV} >M_{GUT}>\sqrt{35\al_{GUT}}\( 6.9\times 10^{15} \){\rm GeV}~.
\eeqa
In our numerical study we input the central values of $g_1$ and $g_2$
while for $g_3$ we require it in the $3\sigma$ at the electroweak scale.
Other inputs at the electroweak scale, for example, the top Yukawa coupling $y_t$,
are extracted from the SM taking into account the threshold corrections.
Relevant details can be seen in the appendix of \cite{FWY,bernal}.
Because of the uncertainty of the GUT scale threshold contributions, we adopt the criteria
that the gauge coupling unification is satisfied when the three couplings differ
within the range  $<0.005$.

\begin{figure}[htbp]
\begin{center}
\includegraphics[width=3.5in]{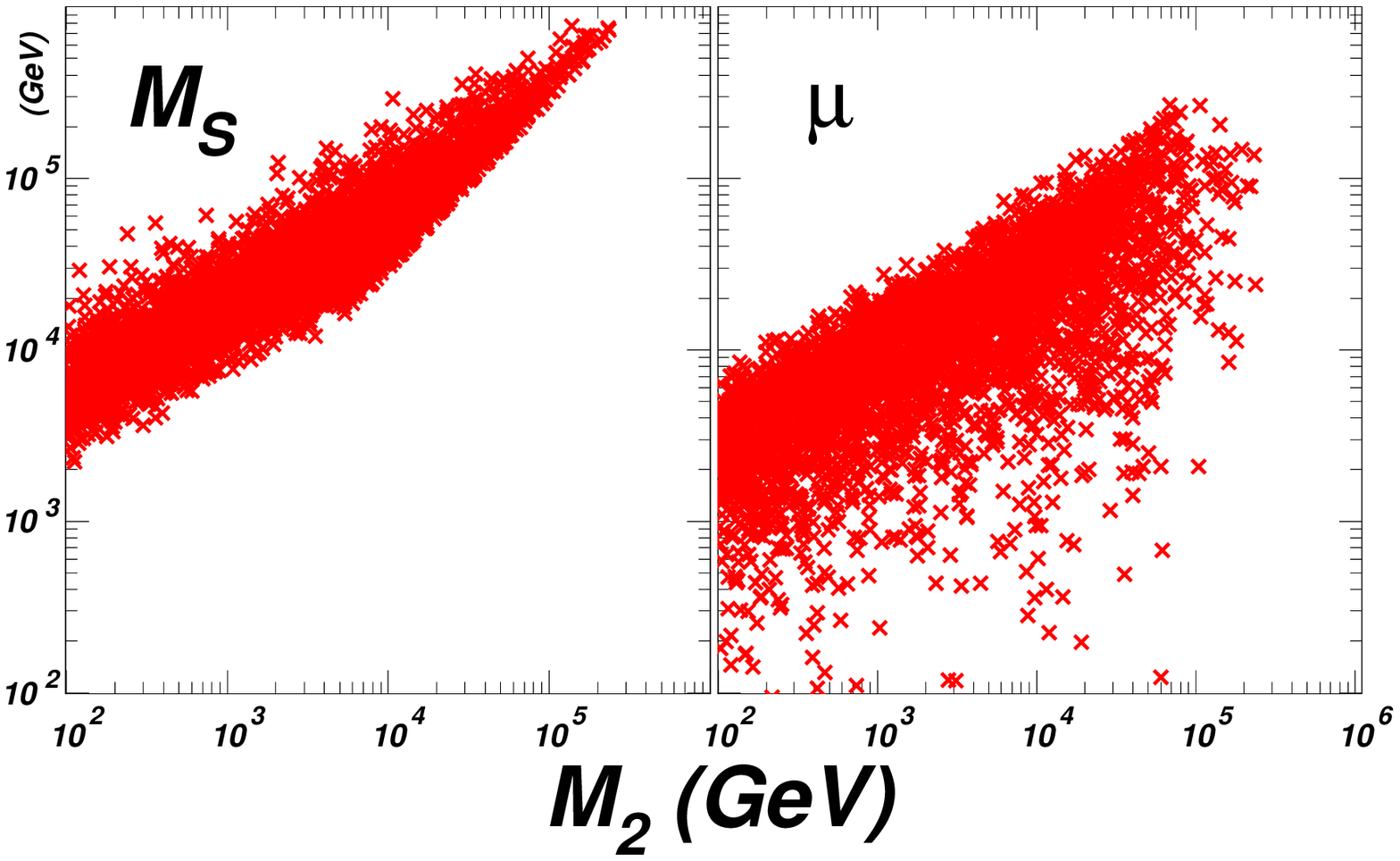}
\includegraphics[width=3.5in]{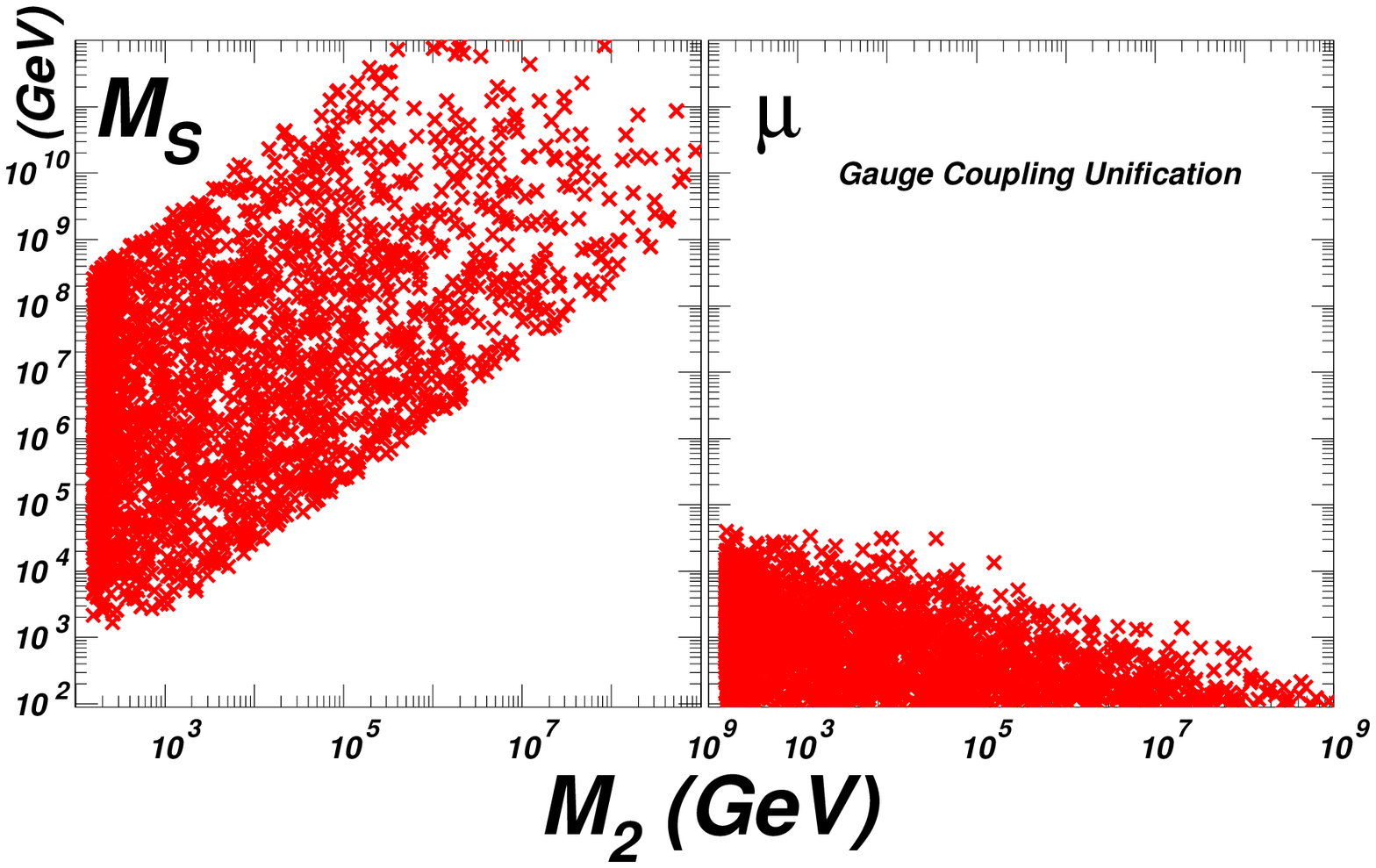}
\end{center}
\vspace{-0.7cm}
\caption{The scatter plots of the parameter space with gauge coupling unification
in case of universal gaugino input (upper) and non-universal gaugino input (lower)
at the GUT scale.}
\label{fig1}
\end{figure}
Fig.\ref{fig1} shows the result of the
parameter space ($M_2$, $M_S$, $\mu$)  with successful gauge coupling unification
 in case of universal condition Eq.(\ref{universal-ew})
and non-universal condition Eq.(\ref{nonuni-ew}), respectively.
We can see that the gaugino unification gives a stringent constraint
on the parameter space. From the upper-left panel we can find
an upper bound for $M_S$, which is about $10^6$ GeV. Since split SUSY requires
 $M_S\gg M_{\tl{g}_i}$, we can also obtain an upper bound for $M_2$ correspondingly.
From the upper-right panel we can find upper limits for $\mu$ and $M_2$, which are
around 100 TeV, independent of the $M_S$ value.
However, the constraints for the non-universal gaugino scenario are rather mild.
From the lower panels, we can see that $M_S$ can be as high as $10^{12}$ GeV
while $M_2$ can be $10^{9}$ GeV.
The $\mu$ parameter, which plays an important role in gauge coupling unification,
also has an upper bound around $10^5$ GeV for both non-universal and universal case.

\subsection{Dark matter constraints}
Now we study the dark matter constraints in our scenario, using the latest dark
matter relic density data from Planck \cite{planck}, WMAP\cite{wmap}
and the direct detection limits from XENON100 \cite{xenon} and the LUX \cite{lux}.
The package DarkSUSY \cite{darksusy} is used to scan the parameter space of
our scenario. In addition to the inputs defined above, we also require
$1<\tan\beta<50$. In order to calculate the relic density of dark matter,
we use the fact that the effects of heavy sfermions and heavy Higgs fields
almost entirely decouple when $M_S=M_A>5 {\rm TeV}$ \cite{FWY0}.
 So in our numerical study, we keep the samples which satisfy the GUT constraints
 and then set $M_S=M_A=10 {\rm ~TeV}$ in DarkSUSY to carry out the dark matter related
 calculations.

For illustration we set $M_S=30,~50,~100,~200~\rm TeV$
and show the allowed ranges for the parameters $M_2, \mu, \tan\beta$.
In our scan, we take into account the current dark matter and collider
constraints (the details can be found in our previous work \cite{FWY}):
\bit
\item[(1)] We require that the relic density of the neutralino
dark matter satisfies the Planck result $\Omega_{DM} = 0.1199\pm 0.0027$ \cite{planck}
(in combination with the WMAP data \cite{wmap}).

\item[(2)] The LEP lower bounds on neutralino and charginos,
including the invisible decay of $Z$-boson.
For LEP experiments, the most stringent constraints come from the
 chargino mass and the invisible $Z$-boson decay.
We require $m_{\tl{\chi}^\pm}> 103 {\rm GeV}$ and the invisible decay width
$\Gamma(Z\ra \tl{\chi}_0\tl{\chi}_0)<1.71~{\rm MeV}$,
 which is consistent with the $2\sigma$ precision EW measurement
 $\Gamma^{non-SM}_{inv}< 2.0~{\rm MeV}$.

\item[(3)] The precision electroweak measurements.
We require the oblique parameters \cite{oblique} to be compatible with
the LEP/SLD data at 2$\sigma$ confidence level \cite{stuconstraints}.

\item[(4)] The combined mass range for the Higgs boson:
$123 {\rm GeV}<M_h <127 {\rm GeV}$ from ATLAS and CMS collaborations of LHC \cite{atlas,cms}.
 In split SUSY, due to large $M_S$, $\log({m_{\tl{f}}^2/m_t^2})\gg 1$ will spoil the convergence of the traditional loop expansion in
evaluating the SUSY effects of Higgs boson self-energy. So in order to calculate mass of the SM-like Higgs boson,
we use the RGE improved effective potential\cite{effpotential} which is employed in the NMSSMTools package\cite{nmssmtools}
after we set $\la=\ka\ra 0$.
\eit
Note that the spin-independent (SI) dark matter-nucleon scattering rate
is calculated with relevant parameters chosen as \cite{Djoudi,Carena,Hisano:2010ct}:
$f_{T_u}^{(p)} =0.023,f_{T_d}^{(p)} = 0.032$,$f_{T_u}^{(n)} = 0.017, f_{T_d}^{(n)} = 0.041$
 and $f_{T_s}^{(p)} = f_{T_s}^{(n)} = 0.020$.
 We take into account all the contributions known so far (including QCD corrections)
in our calculation of the scattering rate.
The value of  $f_{T_s}$  is taken from the recent lattice simulation results \cite{lattice}.

\begin{figure}[htbp]
  \begin{minipage}[t]{0.5\linewidth}
    \centering
    \includegraphics[width=3in]{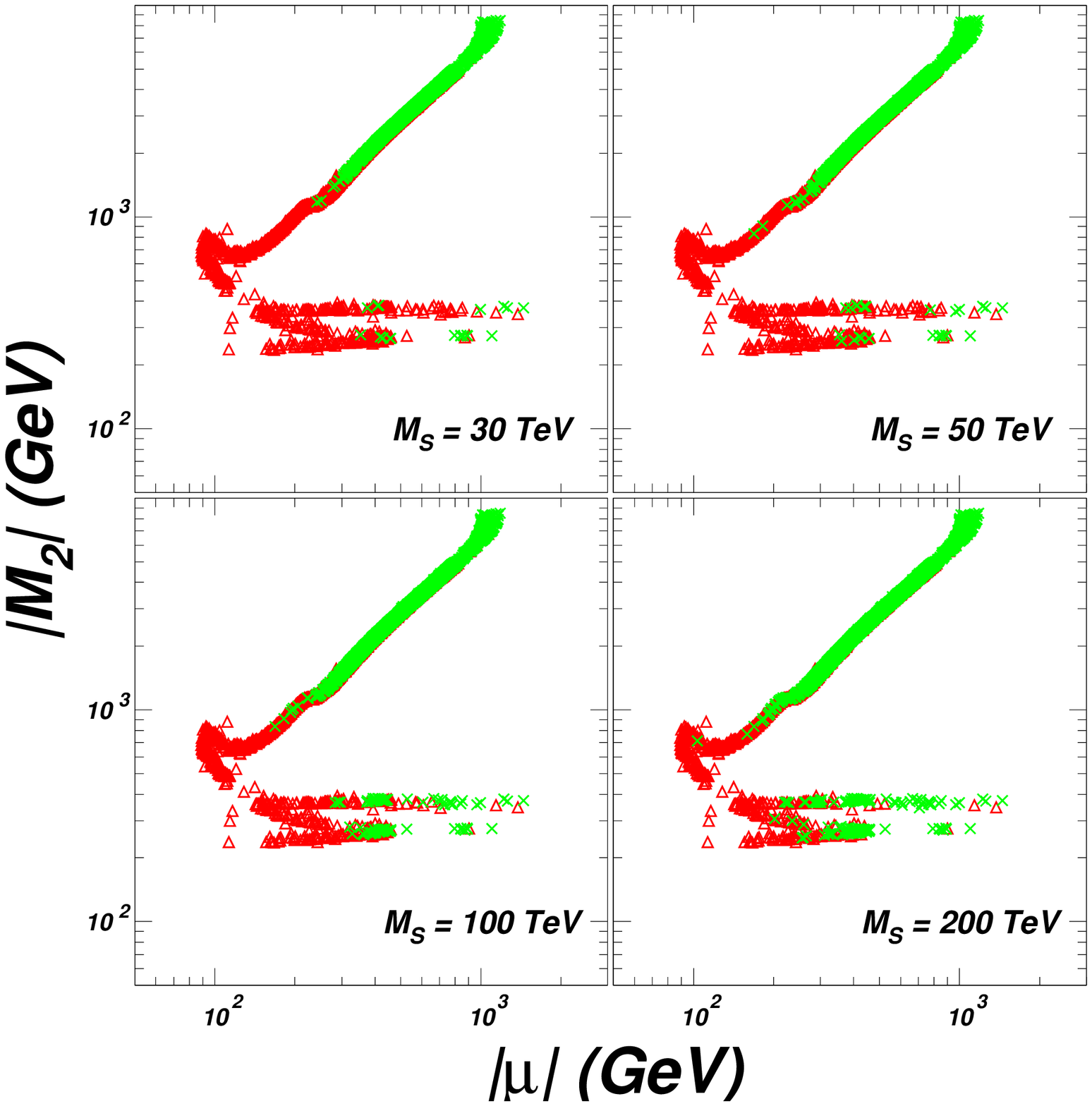}
  \end{minipage}
  \begin{minipage}[t]{0.5\linewidth}
    \centering
    \includegraphics[width=3in]{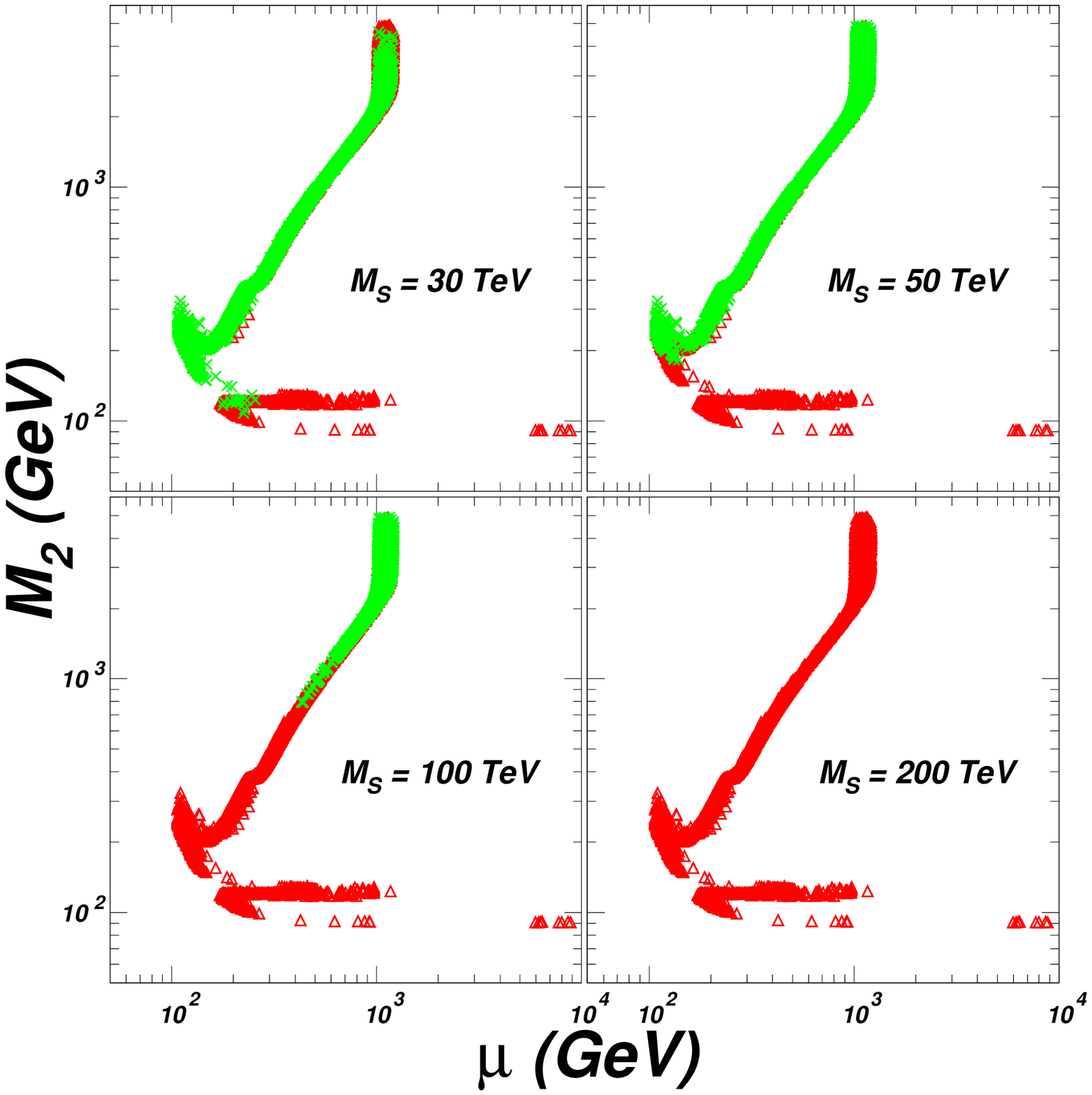}
  \end{minipage}
\caption{The scatter plots of the $(\mu,~M_2)$ parameter space satisfying
constraints (1-4) including dark matter relic density.
The green '$\times$' (red '$\triangle$') can (cannot) achieve the gauge
coupling unification.
The left panel is for the non-universal gaugino scenario proposed in this work
while the right panel is for the universal gaugino scenario studied in our previous work \cite{FWY}.}
\label{fig2}
\end{figure}

\begin{figure}[htbp]
  \begin{minipage}[t]{0.5\linewidth}
    \centering
    \includegraphics[width=3in]{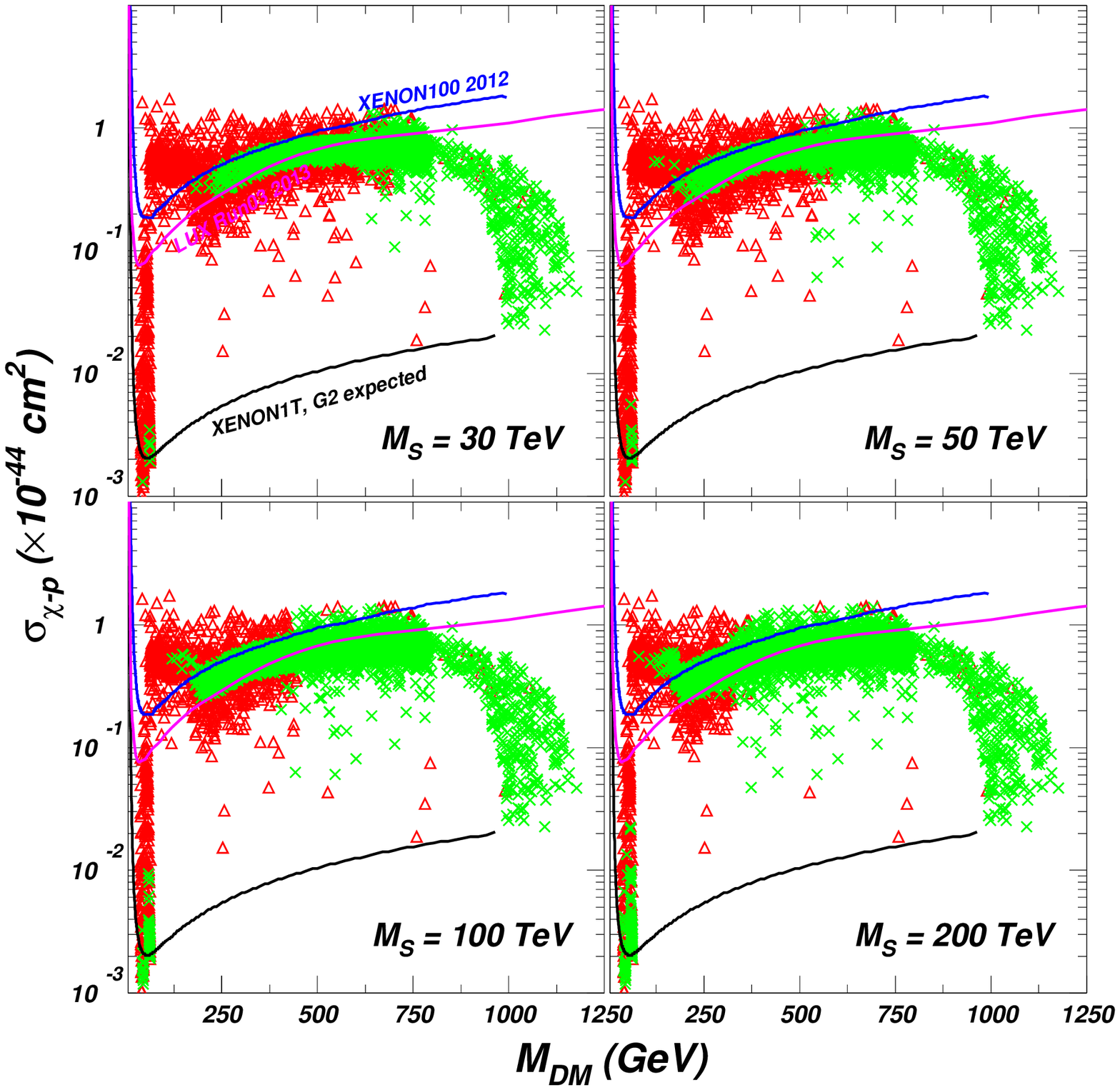}
  \end{minipage}
  \begin{minipage}[t]{0.5\linewidth}
    \centering
    \includegraphics[width=3in]{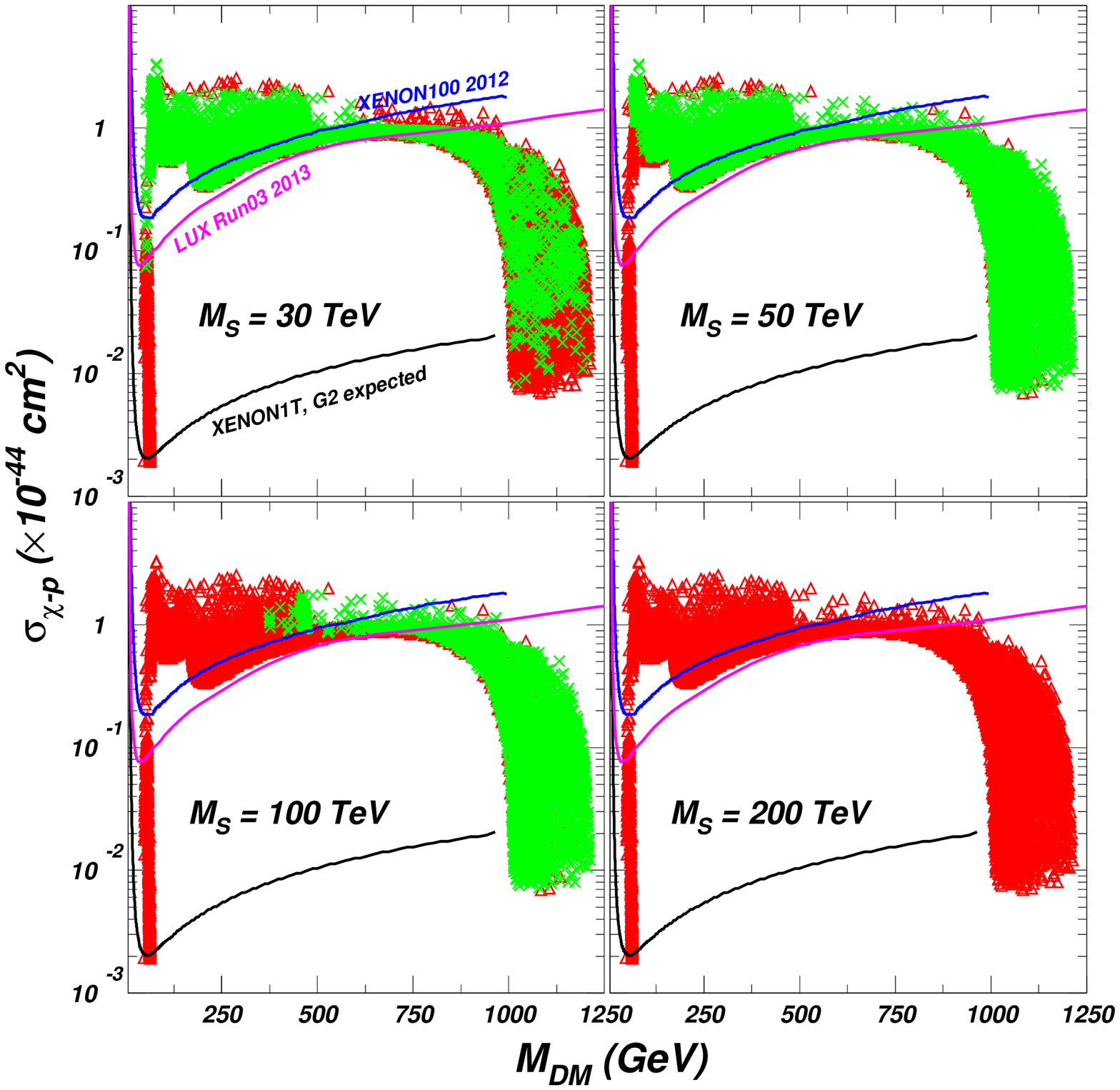}
  \end{minipage}
\caption{Same as Fig.2, but showing the spin-independent cross section
of dark matter scattering off the nucleon.
The curves denote the limits from XENON100(2012) and LUX as well as the future XENON1T sensitivity.
}
\label{fig3}
\end{figure}

It is instructive to compare the dark matter constraints
for non-universal and universal gaugino scenarios.
 Results for the universal gaugino scenario are taken from our previous work \cite{FWY}.
In Figs.\ref{fig2} and \ref{fig3} we show the samples
surviving the constraints (1-4), where the green '$\times$' and
red '$\triangle$' denote respectively the samples allowed and excluded by
the gauge coupling unification requirement. For these results we have the following
discussions:
\bit
\item  We can see from Fig.\ref{fig2} that increasing $M_S$ tends to slightly relax the
gauge coupling unification constraints in non-universal gaugino scenario,
in contrast to universal gaugino scenario where increasing $M_S$ tends to spoil the
gauge coupling unification.

 \item  For both scenarios, a strip corresponding to the higgsino dark matter
with mass range from 1.0 to 1.3 TeV can always survive the combined constraints of
gauge coupling unification and dark matter direct detection (except $M_S>200 {\rm TeV}$
in universal gaugino scenario which is not preferred by gauge coupling unification requirement).
 This is the well known fact that higgsino at about 1.2 TeV can be a viable dark matter candidate.

 \item  Outside the strip of higgsino dark matter,
almost all the survived points will be covered by XEON1T in both scenarios.
 An interesting exception occurs in non-universal gaugino scenario,
where a tiny strip at about 50 GeV cannot be covered by  XEON1T
and such a strip enlarges as $M_S$ increases.
However, a careful analysis indicates that this strip corresponds to a bino dark matter.
 Although the gaugino mass ratio $M_1:M_2:M_3=1:6:12$ is no longer valid at the weak scale
in split SUSY, the RGE running in general
 will not change significantly the mass ratio for a not too large $M_S$.
 So we can estimate that this strip corresponds to a gluino below 700 GeV.
The current preliminary limits on gluino mass using 20 fb$^{-1}$ of 8 TeV data
are $M_{\tl{g}} = 1350$ GeV
(\cite{ATLAS1}) and $M_{\tl{g}} = 1200$ GeV (\cite{CMS1}) assuming a massless neutralino
for mini-split SUSY \cite{gluino-split}.
So this tiny strip should have been ruled out by the LHC data.
 \eit

So we can see that a neutralino dark matter below about 1.0 TeV will be fully covered
by XENON1T for both non-universal and universal gaugino scenarios in split SUSY.
We checked that our numerical calculation results are not sensitive to the sign of $\mu$,
except that for the minus sign a very small part of parameter
space (called blind spots) can survive all the constraints including the future
XENON1T limits \cite{ruderman}.

\section{\label{conclusion}Conclusion}
In this work we proposed to achieve the hierarchy of sparticle spectrum in
split SUSY from the gap between the GUT scale and
the Planck scale.
We built a split SUSY model (which gives non-universal gaugino masses) with proper
high dimensional operators in the framework of SO(10) GUT.
Based on a calculation of two-loop beta functions for gauge couplings
(taking into account all weak scale threshold corrections), we checked the gauge coupling unification
and dark matter constraints.
We found that our scenario can achieve the gauge coupling unification and satisfy
the dark matter constraints in some part of parameter space.
We also examined the sensitivity of the future XENON1T experiment
and found that the currently allowed parameter
space in our scenario
can be covered for a neutralino dark matter below about 1.0 TeV.

\section*{Acknowledgement}
We are very grateful to the referee for useful comments.
This work was supported by the
Natural Science Foundation of China under grant numbers 11105124, 11105125,
11275245, 10821504, 11135003, 11375001, 11172008 and Ri-Xin Foundation of BJUT.

\appendix
\section{The two-loop beta function for gauge couplings}
  In this work, we adopt the method in \cite{twoloop1,twoloop2} to
calculate the two-loop beta functions for three gauge couplings in the split SUSY,
taking into account the weak scale threshold corrections.
Our result is in agreement with \cite{giudice2,giudice1}.
The notations for the two-loop RGE can be found in the appendix of our previous
work \cite{FWY}. The one-loop beta functions for the three gauge couplings are given by
\beqa
b_3&=&-7\theta(\mu-M_Z)+2\theta(\mu-M_{\tl{g}})+2\theta(\mu-M_S)~,\\
b_2&=&-\f{19}{6}\theta(\mu-M_Z)+\f{4}{3}\theta(\mu-M_{\tl{W}})+\f{2}{3}\theta(\mu-M_{\tl{H}})+\f{13}{6}\theta(\mu-M_S)~,\\
b_1&=&\f{41}{10}\theta(\mu-M_Z)+\f{2}{5}\theta(\mu-M_{\tl{H}})+\f{21}{10}\theta(\mu-M_S),
\eeqa
with the step function defined as
\beqa
\theta(x)=\left\{\bea{c}~1,~~~~~x\geq 0;\\~0,~~~~~x<0.\eea\right.
\eeqa

The 2-loop RGE for $SU(3)_c, SU(2)_L, U(1)_Y$ gauge couplings ($g_3,g_2,g_1$, respectively) are given by
\small
\beqa
\f{d}{d\ln E} g_i=\f{b_i}{(4\pi)^2}g_i^3+\f{g_i^3}{(4\pi)^4}\[\sum\limits_{j} B_{ij}g_j^2-\sum\limits_{a=u,d,e}d_i^a Tr(h^{a\da} h^a)-d_W(\tl{g}_u^2+\tl{g}_d^2)-d_B(\tl{g}_u^{\pr 2}+\tl{g}_d^{\pr 2})\]
\eeqa
\normalsize
with the $U(1)_Y$ normalization $g_1^2=\f{5}{3}(g_Y)^2$.

The two-loop beta functions for gauge couplings are given by
\beqa
 B_{ij}&=&\theta(\mu-M_Z)\(\bea{ccc}\f{199}{50}&\f{27}{10}&\f{44}{5}\\\f{9}{10}&\f{35}{6}&12\\\f{11}{10}&\f{9}{2}&-26\eea\)+\theta(\mu-M_2)\(\bea{ccc}0&0&0\\0&\f{64}{3}&0\\0&0&0\eea\)+\theta(\mu-M_3)\(\bea{ccc}0&0&0\\0&0&0\\0&0&48\eea\)
\nn\\&+&\theta(\mu-M_{\tl{H}})\(\bea{ccc}\f{9}{50}&\f{9}{10}&0\\\f{3}{10}&\f{49}{6}&0\\0&0&0\eea\)+\theta(\mu-M_S)\(\bea{ccc}\f{19}{5}&\f{9}{5}&\f{44}{5}\\\f{3}{5}&-\f{31}{3}&12\\\f{11}{10}&\f{9}{2}&-8\eea\).
\eeqa
Similarly, we have
\beqa
d^u&=&\theta(\mu-M_Z)(\f{17}{10},\f{3}{2},2)+\theta(\mu-M_S)(\f{7}{2},\f{9}{2},2)~,\\
d^d&=&\theta(\mu-M_Z)(\f{1}{2},\f{3}{2},2)+\theta(\mu-M_S)(\f{23}{10},\f{9}{2},2)~,\\
d^e&=&\theta(\mu-M_Z)(\f{3}{2},\f{1}{2},0)+\theta(\mu-M_S)(\f{21}{10},\f{3}{2},0)~,
\eeqa
\beqa
d^W&=&(\f{9}{20},\f{11}{4},0)\theta(\mu-\max(M_2,M_{\tl{H}}))+\theta(\mu-M_S)(-\f{9}{20},-\f{11}{4},0)~,\\
d^B&=&(\f{3}{20},\f{1}{4},0)\theta(\mu-\max(M_1,M_{\tl{H}}))+\theta(\mu-M_S)(-\f{9}{20},-\f{11}{4},0)~.
\eeqa
 The one-loop renormalization group equations for Yukawa couplings below the $M_S$ scale
can be written as
\beqa
16\pi^2 \f{d}{dt} h^u&=& h^u\[-3c_i^u g_i^2+c_{T}^u T+ c_{S_1}^u S_1+ c_{S_2}^u S_2+\f{3}{2}\(h^{u\da}h^u-h^{d\da}h^d\)\],\\
16\pi^2 \f{d}{dt} h^d&=& h^d\[-3c_i^d g_i^2+c_{T}^d T+c_{S_1}^d S_1+ c_{S_2}^d S_2+\f{3}{2}\(h^{d\da}h^d-h^{u\da}h^u\)\],\\
16\pi^2 \f{d}{dt} h^e&=& h^e\[-3c_i^e g_i^2+c_{T}^e TT+c_{S_1}^e S_1+c_{S_2}^e S_2+\f{3}{2}h^{e\da}h^e\],
\eeqa
with
\beqa
T&=&Tr(3h^{u\da}h^u+3h^{d\da}h^d+h^{e\da}h^e),~S_1=\f{1}{2}\[(\tl{g}_u^\pr)^2+(\tl{g}_d^\pr)^2\],~S_2=\f{3}{2}\(\tl{g}_u^2+\tl{g}_d^2\).
\eeqa
Above $M_S$, we will recover the MSSM result and the one-loop RGE for Yukawa-type interactions
in the superpotential are well known to be
\beqa
16\pi^2 \f{d}{dt} \la^u&=& \la^u\[-2c_i^u g_i^2+3Tr(\la^{u\da}\la^u)+3\la^{u\da}\la^u+\la^{d\da}\la^d\],\\
16\pi^2 \f{d}{dt} \la^d&=& \la^d\[-2c_i^d g_i^2+Tr(3\la^{d\da}\la^d+\la^{e\da}\la^e)+\la^{u\da}\la^u+3\la^{d\da}\la^d\],\\
16\pi^2 \f{d}{dt} \la^e&=& \la^e\[-2c_i^e g_i^2+Tr(3\la^{d\da}\la^d+\la^{e\da}\la^e)+3\la^{e\da}\la^e\],
\eeqa
with
\beqa
c_i^u=(\f{13}{30},\f{3}{2},\f{8}{3}),~c_i^d=(\f{7}{30},\f{3}{2},\f{8}{3}),~c_i^e=(\f{9}{10},\f{3}{2},0).
\eeqa
The one-loop Yukawa couplings for $c_i^u,c_i^d,c_i^e$ are calculated to be
\beqa
\(\bea{c}
c_i^u\\c_i^d\\c_i^e
\eea\)=\theta(\mu-M_Z)\(\bea{ccc}~\f{17}{60}&\f{3}{4}&\f{8}{3}\\\f{1}{12}&\f{3}{4}&\f{8}{3}\\\f{3}{4}&\f{3}{4}& 0\eea\)
+\theta(\mu-M_S)\(\bea{ccc}~\f{3}{20}&\f{3}{4}&0\\\f{3}{20}&\f{3}{4}&0\\\f{3}{20}&\f{3}{4}& 0\eea\)\eeqa
All terms are set to zero above $M_S$. Besides, we have
\beqa
\(\bea{ccc} c^u_{T}&c^u_{S_1}&c^u_{S_2}\\ c^d_{T}&c^d_{S_1}&c^d_{S_1}\\ c^e_{T}&c^e_{S_1}&c^e_{S_2}\eea\)&=&
\theta(\mu-M_Z)\(\bea{ccc}1&0&0\\1&0&0\\1&0&0\eea\)+\theta(\mu-\max(M_2,M_{\tl{H}}))\(\bea{ccc}0&0&1\\0&0&1\\0&0&1\eea\)~,\nn\\
&+&\theta(\mu-\max(M_1,M_{\tl{H}}))\(\bea{ccc}0&1&0\\0&1&0\\0&1&0\eea\)
\eeqa
and all set to zero above $M_S$.

One loop RGE for gaugino couplings $\tl{g}_u,\tl{g}_u^\pr,\tl{g}_d,\tl{g}_d^\pr$ below $M_S$
are given as (with the gaugino relation $M_1<M_2$)
\bit
\item Between $[\max(M_2,M_{\tl{H}}),M_S]$, the RGE for $\tl{g}_{u,d}$ are given by
\scriptsize
\beqa
\label{aa}
16\pi^2\f{d }{dt} \tl{g}_{u}&=&-3\tl{g}_u c_i^u g_i^2+\f{5}{4}\tl{g}_u^3-\f{1}{2}\tl{g}_u\tl{g}_d^2+\f{1}{4}\tl{g}_u\tl{g}_u^{\pr 2}+\tl{g}_d\tl{g}_d^\pr\tl{g}_u^\pr+\tl{g}_u (T+c_{S_1} S_1+c_{S_2}S_2)\\
16\pi^2\f{d }{dt} \tl{g}_{d}&=&-3\tl{g}_d c_i^d g_i^2+\f{5}{4}\tl{g}_d^3-\f{1}{2}\tl{g}_d\tl{g}_u^2+\f{1}{4}\tl{g}_d\tl{g}_d^{\pr 2}+\tl{g}_u\tl{g}_u^\pr\tl{g}_d^\pr+\tl{g}_d (T+c_{S_1} S_1+c_{S_2}S_2)
\eeqa
\normalsize
with the coefficient
\beqa
c_i^{u,d}=(\f{3}{20},\f{11}{4},0),~~c_{S_1}=c_{S_2}=1.
\eeqa
Below $\max(M_2,M_{\tl{H}})$, the coupling are switched off.

\item  Between $[\max(M_2,M_{\tl{H}}),M_S]$, the RGE for $\tl{g}_{u,d}$ are given by
\scriptsize
\beqa
\label{bb}
16\pi^2\f{d }{dt} \tl{g}_{u}^\pr&=&-3\tl{g}_u^\pr \tl{c}_i^u g_i^2+\f{3}{4}\tl{g}_u^{\pr 3}+\f{3}{2}\tl{g}^\pr_u\tl{g}_d^{\pr 2}+\f{3}{4}\tl{g}_u^\pr\tl{g}_u^{2}+3\tl{g}_d^\pr\tl{g}_d\tl{g}_u+\tl{g}_u^\pr (T+c_{S_1} S_1+c_{S_2}S_2) \\
16\pi^2\f{d }{dt} \tl{g}_{d}^\pr&=&-3\tl{g}_d^\pr \tl{c}_i^d g_i^2+\f{3}{4}\tl{g}_d^{\pr 3}+\f{3}{2}\tl{g}^{\pr}_ d\tl{g}^{\pr 2}_u+\f{3}{4}\tl{g}^{\pr}_d
\tl{g}_d^{2}+3\tl{g}_u^\pr\tl{g}_u \tl{g}_d+\tl{g}_d^\pr (T+c_{S_1} S_1+c_{S_2}S_2)
\eeqa
\normalsize
with the coefficient
\beqa
\tl{c}_i^{u,d}=(\f{3}{20},\f{3}{4},0),~~c_{S_1}=c_{S_2}=1.
\eeqa
Between $[\max(M_1,M_{\tl{H}}),\max(M_2,M_{\tl{H}})]$, the RGE reads
\beqa
\label{cc}
16\pi^2\f{d }{dt} \tl{g}_{u}^\pr&=&-3\tl{g}_u^\pr \tl{c}_i^u g_i^2+\f{3}{4}\tl{g}_u^{\pr 3}+\f{3}{2}\tl{g}^\pr_u\tl{g}_d^{\pr 2}+\tl{g}_u^\pr (T+c_{S_1} S_1+c_{S_2}S_2),\\
16\pi^2\f{d }{dt} \tl{g}_{d}^\pr&=&-3\tl{g}_d^\pr \tl{c}_i^d g_i^2+\f{3}{4}\tl{g}_d^{\pr 3}+\f{3}{2}\tl{g}^{\pr}_d\tl{g}^{\pr 2}_u+\tl{g}_d^\pr (T+c_{S_1} S_1+c_{S_2}S_2),
\eeqa
with the coefficient
\beqa
\tl{c}_i^{u,d}=(\f{3}{20},\f{3}{4},0),~~c_{S_1}=1,~~c_{S_2}=0.
\eeqa
Below $\max(M_1,M_{\tl{H}})$, the coupling are switched off.
\eit

\end{document}